# CONSIDERATIONS FOR A DEDICATED GEONEUTRINO DETECTOR FOR GEOSCIENCES

## P. ILA[1], W. GOSNOLD[2], P. JAGAM[3], G. I. LYKKEN[4]

1. Department of Earth, Atmospheric and Planetary Sciences,
   Massachusetts Institute of Technology, Cambridge, MA 02139
   The NORM Group Organization, Cambridge, MA, USA 02139
   pila@mit.edu; ila@thenormgroup.org

2. Department of Geology and Geological Engineering,
   University of North Dakota, Grand Forks, ND, USA 58202
   willgosnold@mail.und.edu

3. Department of Physics,
   University of Guelph, Guelph, ON, Canada N1G1L9
   SNO Collaboration, Canada;
   The NORM Group Organization, Guelph, ON, Canada N1G1L9
   pjagam@uoguelph.ca; jagam@thenormgroup.ca

4. Department of Physics,
   University of North Dakota, Grand Forks, ND, USA 58202
   glenn_lykken@und.nodak.edu

## For the GRAFG collaboration

[GRAFG – Geoneutrino Radiometric Analysis For Geosciences]

**Contact info:**
**pila@mit.edu**
**ila@thenormgroup.org**

# ABSTRACT

A combination of several sources including: radiogenic heating, processes of mantle and core formation and differentiation, delayed radiogenic heating, earthquakes, and tidal friction account for the surface heat flux in the Earth. Radiogenic heating is of much interest in various fields of geosciences. Inferences from recent experiments with reactor antineutrinos and solar neutrinos showed that the age of geoneutrinos is at hand for constraining radiogenic heat. Because of the deep penetrating properties of the neutrinos this type of radiation in the decay of the heat producing elements (HPE) is ideally suited for the investigation of the deep interiors of the Earth compared to conventional radiometric methods for HPE employing alpha-, beta- and gamma rays. This presentation will address the considerations for a dedicated geoneutrino detector to be set up for investigating the interior regions all the way to the center of the Earth.

# INTRODUCTION

Knowledge of energy sources in the Earth is of increasing interest from many different points of view in the geosciences. Using geothermal conductivity information Kelvin estimated the age of the Earth. This age estimate was not acceptable when compared to other evidence. This disagreement led to the searches for the identification of other sources of heat production within the Earth.

## Natural radioactivity as a heat source in the Earth and Heat Producing Elements (HPE):

Natural radioactivity in the Earth was quickly recognized as a heat source soon after its discovery. Radiogenic heat was investigated as the source of heat flux over and above the primordial heat in the Earth. Abundances of the heat producing elements in the Earth, namely K, U, Th, in the Earth's crust are investigated extensively.

## Dependence of investigations for assayingthe HPE:

These investigations were dependent on geological sampling and geochemical assay techniques. Radiometric and X-ray techniques for assaying HPE evolved rapidly with developments in instrumental analysis. Techniques based on radioactive radiations exploited the signals generated by characteristic $\alpha-$, $\beta-$, $\gamma-$ and x-rays from the HPE with high resolution and high sensitivity radiation detectors.

## Limitations of geochemical assay techniques and sampling:

These instruments and techniques were limited by the penetrating power of the radiation used in the assay techniques employed in the laboratory, for in-situ assaying in the field, or in the context of assaying the whole Earth. Over a period of time, the limitations of the geochemical assay techniques developed for HPE determinations based on sampling techniques were identified. In-situ sampling was needed to reduce the cost of field sampling. Techniques for sampling at ever increasing depths from the surface were needed to investigate the interior regions of the Earth.

**Particle Physics instrumentation capability:**

Interest in the investigation of the particle physics properties of a weakly interacting highly penetrating radioactive radiation led to the development of sophisticated instruments and sensitive techniques for detecting this radiation. It is these instruments and techniques, which are of interest to the geoscientists for the investigation of the deep interior of the Earth. This presentation tries to discuss and understand the basic considerations in the selection and deployment of these particle physics instruments and techniques for whole Earth assay of HPE, and for tomographic assay to assign the HPE to specific regions of the Earth.

**Focus of this presentation:**

The focus of this presentation is to

  1) identify the considerations for the selection and deployment of dedicated instruments for detecting the HPE concentrations in the deep interiors of the Earth, which are otherwise inaccessible by conventional sampling techniques,

  2) discuss and understand the basic considerations in the selection and deployment of these particle physics instruments and techniques for whole Earth assay of HPE and for tomographic assay to assign the HPE to specific regions of the Earth.

# NEUTRINO TERMINOLOGY AND PROPERTIES.

## Terminology:

Neutrinos $(\nu)$ and antineutrinos $(\overline{\nu})$ are emitted in nuclear positive and negative $\beta$ decays. Together they are generally called neutrinos except when referring to the specific type. An example of neutrino emission from a HPE is shown below in the decay of Potassium-40.

**Positive $\beta$ decay**

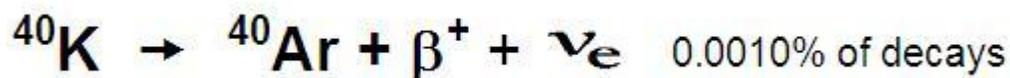
$^{40}K \rightarrow {}^{40}Ar + \beta^+ + \nu_e$    0.0010% of decays

**Negative $\beta$ decay**

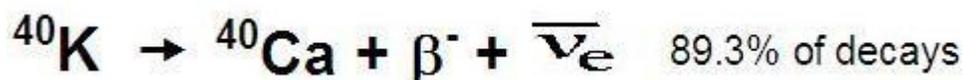
$^{40}K \rightarrow {}^{40}Ca + \beta^- + \overline{\nu}_e$    89.3% of decays

The subscript on the symbol for neutrino indicates that the neutrinos are emitted in nuclear beta-decay compared to other types of neutrinos emitted in other types of radioactive decays, which emit other types of neutrinos.

**Positive beta decay accompanied by neutrinos:**
A proton in an unstable nucleus becomes a neutron emitting a positron and a neutrino.

$$p \rightarrow n + \beta^+ + \nu_e$$

**Negative beta decay accompanied by antineutrinos:**
A neutron in an unstable nucleus becomes a proton emitting an electron and an antineutrino.

$$n \rightarrow p + \beta^- + \overline{\nu_e}$$

**Inverse beta decay capturing antineutrinos by protons in a detector:**
An incoming antineutrino interacts with a proton in the detection medium releasing a neutron and a positron.

$$\overline{\nu_e} + p \rightarrow n + \beta^+$$

**Penetrating power and directionality:**

Compared to the $\alpha-$, $\beta-$, $\gamma-$ and x-rays emitted in the radioactive decay of unstable elements, neutrinos are weakly interacting particles. Therefore, they are not stopped or scattered from their initial direction of travel or, their intensity attenuated by absorbers or shielding materials commonly used with the other radiations as shown in figure 1.

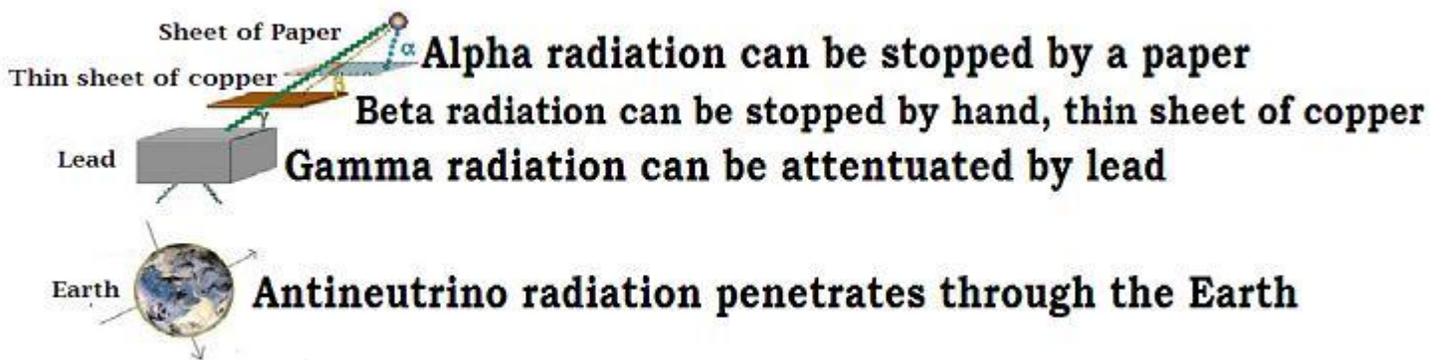

**Figure 1**. Relative advantage of using antineutrinos as a radiation probe in radiometric techniques for the assay of the heat producing elements (HPE). Neutrinos travel in straight lines from the point of origin to the point of detection. This characteristic is advantageous for bulk in-situ assay in the field or, for whole Earth assay and for tomography of the radioactivity in the Earth.

# GEONEUTRINOS AND HEAT PRODUCING ELEMENTS

Geoneutrinos are the antineutrinos produced during the negative beta-decay of the long-lived isotopes of the unstable elements in the Earth. The predominant production is from Potassium, Uranium, and Thorium, which are usually referred to as radiogenic Heat Producing Elements (HPE). The antineutrino production rates for the HPE are given in Table 1.

**Table 1.** Radioactive half-lives and antineutrino production rates of the predominant heat producing elements in the Earth [Ref. Eder].

| HPE | Half Life Billion years | Antineutrino Production rate $kg^{-1} sec^{-1}$ |
|---|---|---|
| U-238 | 4.47 | $2.69 \times 10^4$ |
| Th-232 | 14.00 | $1.63 \times 10^7$ |
| K-40 | 1.28 | $7.41 \times 10^7$ |
| | | based on isotopic abundance |

The production rates quoted above are estimates from isotopic abundances of elemental uranium, thorium and potassium by weight.

The antineutrino production rates given in Table 2 are calculated from known concentrations of HPE in the specified regions of the Earth. For the lower mantle, observational data are not available; the production rates are based on the Bulk Silicate Earth model that describes the present crust-plus-mantle system based on geochemical arguments. According to geochemical arguments, negligible amounts of U, Th, and K should be present in the core [Ref. Mantovani et al].

**Table 2**. Estimated antineutrino production rates of the predominant heat producing elements in the Earth. Note that the estimate from the core regions of the Earth is zero [Ref. Mantovani et al].

| Earth Shells | Antineutrinos/Second | | |
| --- | --- | --- | --- |
| | K | Th | U |
| Continental Crust | 1.03E+25 | 2.32E+24 | 2.56E+24 |
| Oceanic Crust | 2.07E+23 | 2.08E+22 | 4.32E+22 |
| Upper Mantle | 2.32E+24 | 2.94E+23 | 5.05E+23 |
| Lower Mantle | 1.32E+25 | 2.45E+24 | 2.84E+24 |
| Outer Core | 0 | 0 | 0 |
| Inner Core | 0 | 0 | 0 |

## Antineutrinos of K, U and Th:

From the above table it can be seen that potassium in the Earth is the dominant producer of antineutrinos. Thorium and uranium contribute about the same to the antineutrino production rate. However, the antineutrinos emitted by potassium have lower distribution of energies compared to those emitted from thorium and uranium. This difference in the antineutrino spectra has to be taken into account when considering the detection of HPE by instrumental techniques. In addition, to identify thorium and uranium separately with antineutrino detectors, consideration should be given to the instrumentation capabilities for resolving the data by energy dispersive spectrometry. These considerations will be elaborated below.

## Earth as a source of antineutrinos:

The different Earth regions are shown schematically in figure 2. It should be recognized that the Earth is a distributed volume source of antineutrinos emitted by the HPE. In addition, the average concentrations in the different regions of the individual HPE are widely different over and above the local variations due to mineralization.

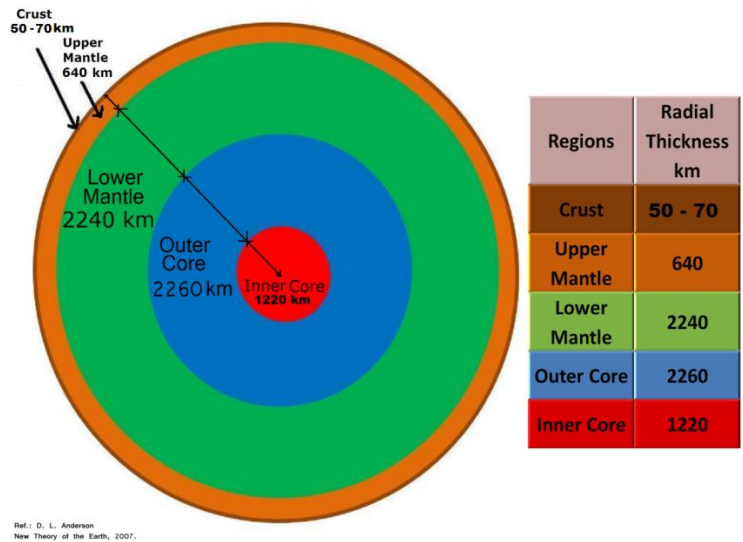

**Figure 2**. Cross-sectional view of the interior regions of the Earth based on Anderson (2007). The radial thicknesses of the different regions are labeled in the figure. Compared to the total radius of 6370 km of the Earth, the deepest drill cores were obtained from a maximum depth of only 10 km from the surface of the Earth. The inner core of the Earth is totally inaccessible for sampling by drilling methods. Volcanoes to some extent provide access to the upper mantle.

# CONSIDERATIONS FOR A DEDICATED GEONEUTRINO DETECTOR FOR GEOSCIENCES

## I. GENERAL CONSIDERATIONS:

Geosciences considerations depend on the research area of interest, for selecting and deploying an optimized antineutrino detector of choice:

- Corroborating and improving the precision of the HPE assays already done: This requires detectors of high sensitivity with spectroscopic capability for doing in-situ assays of HPE.

- Assaying the total radioactivity in the Earth including the core without regard to individual HPE: In contrast to the above consideration, another consideration to be taken into account is assaying the total radioactivity in the Earth including the core without regard to individual HPE. In this case the spectroscopic capability is not required and, a detector with lower detection sensitivity may be acceptable.

- Understanding the discrepancies in heat flow calculations: Another consideration is identifying the concentrations of HPE in the interior regions of the Earth inaccessible by current sampling methods seems in order, to understand the discrepancies in heat flow calculations.

- Overall cost of the optimized detector of choice: The sensitivity and spectroscopic capability requirements determine the overall cost of the optimized detector of choice for the particular application in the research area of interest.

## II. SELECTION CONSIDERATIONS OF ANTINEUTRINO DETECTOR TYPES:

In general, there are two types of antineutrino detectors:
    1) with directional sensitivity,
    2) without directional sensitivity.

If localizing the HPE radioactivity is a consideration when selecting a detector with a particular lower limit of detection sensitivity, then the type of antineutrino detector that will meet this requirement needs to be taken into account.

In this context, it should be pointed out that the scintillation detectors based on the principle of inverse beta decay for the detection of antineutrinos are only capable of giving the average concentrations of the HPE in the whole Earth provided they have the minimum detection sensitivity for the particular application. Because of their lack of directional sensitivity they cannot localize precisely where the activity originates from the HPE of interest.

### III. TOMOGRAPHY CONSIDERATIONS:

Practical detectors have been designed and built with directional sensitivity:
- They are a class of scintillation detectors known as Cérenkov detectors.
- They have been used primarily for studies of energy production in the Sun and astrophysical phenomena such as Super Novae.
- They suffer from limited detection sensitivity for geoscience applications at this time.
- They can be optimized for applications in geosciences to improve the detection sensitivity.
- They require deep underground locations to minimize the cosmic-ray component of the background while maximizing the signal to background ratio in the detection of the signals from the HPE in the Earth.

Together with spectroscopic capability to identify the concentrations of the individual HPE in the different regions of the Earth the Cérenkov detectors are the most expensive and time consuming antineutrino detectors to build and operate. Figure 3 shows conical views of the interior regions of the Earth from the detection point of view requiring tomographic design considerations. Different tomographic designs are under considerations for geoneutrino studies (Ref. De Meijer et al) and core density determination (Ref. Winter).

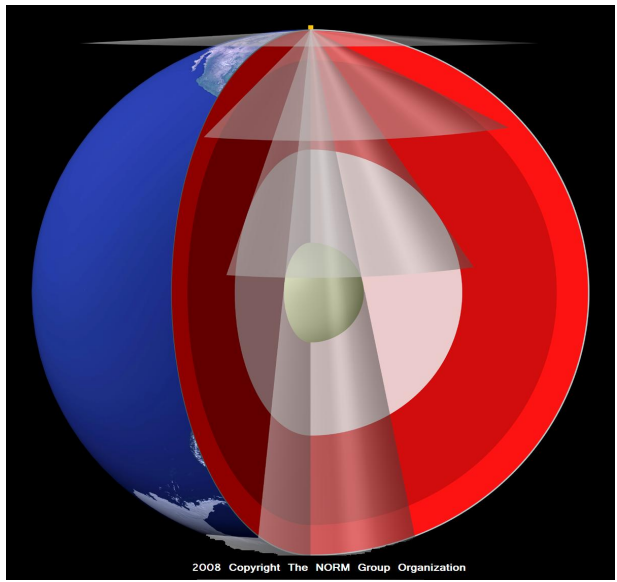

**Figure 3.** Localization of the HPE with directionally sensitive tomographic detectors. The concentration of HPE in the different regions of the Earth can be identified with Cérenkov antineutrino detectors which provide directional and spectroscopic capabilities. (Figure credit: The NORM Group Organization, Cambridge, MA).

## IV. SPECTROSCOPIC CONSIDERATIONS:

A neutron in an unstable nucleus becomes a proton, emitting an electron and an antineutrino.

$$n \rightarrow p + \beta^- + \overline{\nu}_e$$

For example, elemental potassium consists of three isotopes of which $^{40}$K is radioactively unstable and decays with a half-life of 1.28 billion years. Both antineutrinos and neutrinos are emitted in the decay in the following proportions.

$$^{40}K \rightarrow {}^{40}Ca + \beta^- + \overline{\nu}_e \quad 89.3\% \text{ of decays}$$
$$^{40}K \rightarrow {}^{40}Ar + \beta^+ + \nu_e \quad 0.0010\% \text{ of decays}$$

The antineutrinos and neutrinos emitted in the decay of $^{40}$K are not mono-energetic. The spectrum of energies extends from zero to a maximum energy characteristic of the radioactive decay. The nuclear decay scheme of $^{40}$K is shown in figure 4.

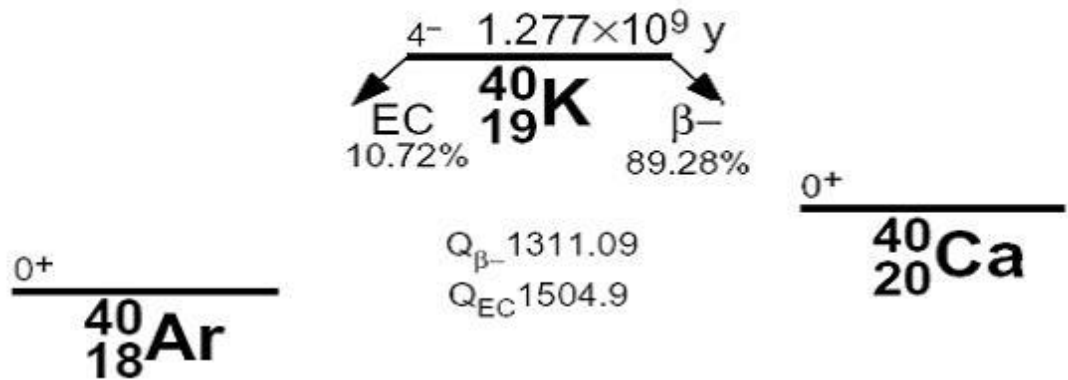

89.28% $Q_\beta$=1.311 MeV

10.72% $Q_{EC}$=1.505 MeV

10.67% to 1.461 MeV state ($E_\nu$ = 44 keV)

0.05% to g.s. ($E_\nu$ = 1.5 MeV)

**Figure 4.** Radioactive decay properties of Potassium – 40.

Ref.: Table of Isotopes, Seventh Edition, Eds. Lederer and Shirley.

Energy distributions in the antineutrino spectra of all the important heat producing elements in the Earth is shown in figure 5. The antineutrino intensity falls off to zero with increasing energy. The vertical dotted line shows the threshold energy for detection by detectors employing inverse beta decay reaction with protons in a hydrogenous detecting medium.

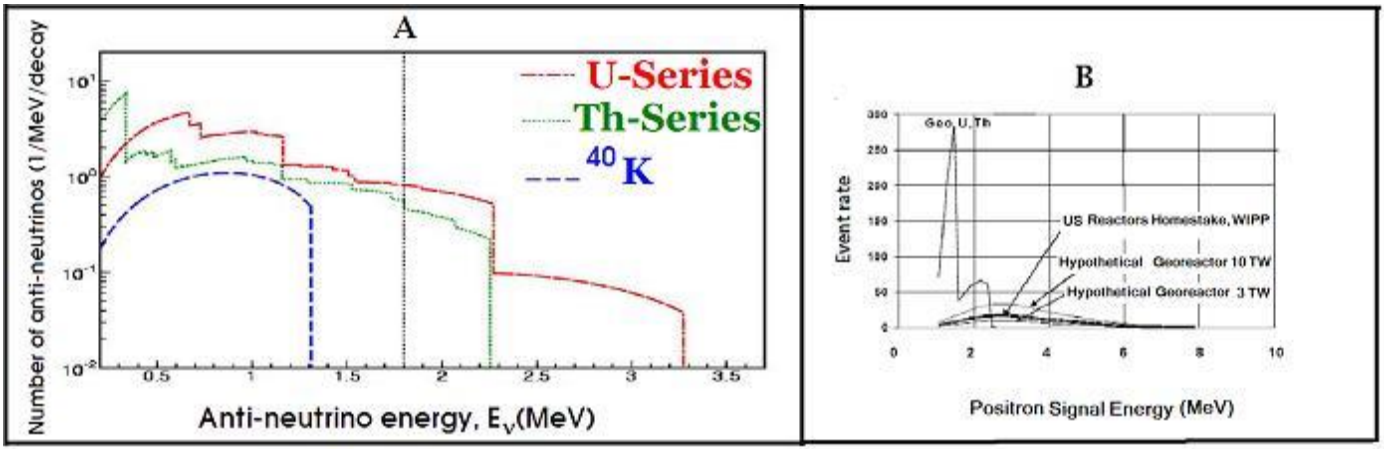

**Figure 5. A**. Spectral distribution of antineutrino energies emitted in the decay of principal HPE in the Earth. The vertical dotted line at 1.8 MeV represents the energy threshold of scintillation detectors employing the inverse beta decay reaction for the detection of antineutrinos. (Ref. KamLAND collaboration: Araki et al.). **B**. The energy spectrum of antineutrinos (geoneutrinos) upto 8 MeV energy (Ref. Raghavan) including fission antineutrinos.

Geoneutrinos (antineutrinos) occurring in the decay of HPE are confined to energies below 3 MeV. The vertical dotted line in figure 5A shows the threshold energy for detection by detectors employing inverse beta decay reaction on protons. Therefore, most of the geoneutrino intensity is below the detection threshold of these detectors. In addition, the inverse beta decay reaction cross-section falls off at lower energies than fission neutrino energies limiting the detection sensitivity of these detectors.

## HPE ASSAYING WITH ANTINEUTRINO DETECTORS

### I. RADIOMETRIC METHOD

Assuming isotropic emission of radiations by a decaying nucleus, the count rate detected in a practical detector is given by

$$\mathbf{N_d = N_0\ \Omega\ \varepsilon}$$

where $N_0$ is the rate of emission per unit time into $4\pi$ space, $\Omega$ is the fractional solid angle and $\varepsilon$ is the intrinsic efficiency of the detector which can be determined experimentally. This is the principle of the comparator method widely used in the radiometric analysis of HPE employing $\alpha-$, $\beta-$, $\gamma-$ and x-rays. The detector is first calibrated for detection efficiency with a known source whose decay rate is known. The unknown source decay rate is determined from the measured counts.

## II. ABSOLUTE METHOD

Alternatively, the count rate detected can be used to determine, **$N_d$,** the flux of radiation going through the detector from the equation

$$\mathbf{N_d} = \Phi \, \mathbf{W} \, \sigma$$

where $\Phi$ is the flux of radiation going through the detector per unit time, **W** is the number of protons in the detector and $\sigma$ is the interaction cross-section per proton. This is the principle of the absolute method.

The interaction cross-section is dependent on the energy of the antineutrino. Inverse beta decay reaction cross-sections with protons are needed in the energy regions of interest in the beta decays of HPE. After calculating the flux from the measured counting rates, the source decay rate can be calculated.

# DISCUSSION OF PRACTICAL ANTINEUTRINO DETECTORS

Practical antineutrino detectors were designed and tested for specific purposes for over fifty years.
- The first successful detector design demonstrated the detection principle by inverse beta decay.
- Improvements based on this design led to the experimental determination of the interaction cross-section for inverse beta decay using the power reactors as anthropogenic antineutrino sources. This cross-section measurement was specific to the fission antineutrino sources and specific to the fission energy region from about 2 MeV to 10 MeV. The lower limit of 2 MeV for the region of antineutrinos detected, arises from the reaction threshold for the inverse beta decay on protons, where as geoneutrinos occur in the decay of HPE are confined to energies below 3 MeV.

A meter cubed antineutrino detector is shown in figure 6 and a spectrum recorded with that detector near a power reactor is shown in figure 7.

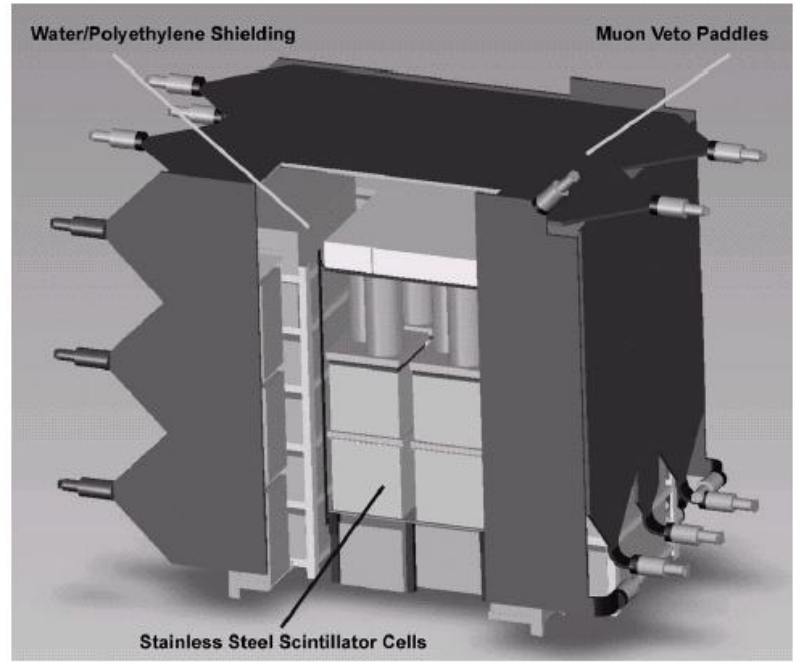

**Figure 6.** Cut away diagram of a meter cubed detector detecting antineutrinos in the vicinity of a power reactor (Ref. Bernstein et al).

Geoneutrinos occur in the decay of HPE and are confined to energies below 3 MeV. Therefore, most of the geoneutrino intensity is below the detection threshold of the detectors employing inverse beta decay reaction. In addition, the inverse beta decay reaction cross-section falls off at lower energies than fission neutrino energies limiting the detection sensitivity of these detectors.

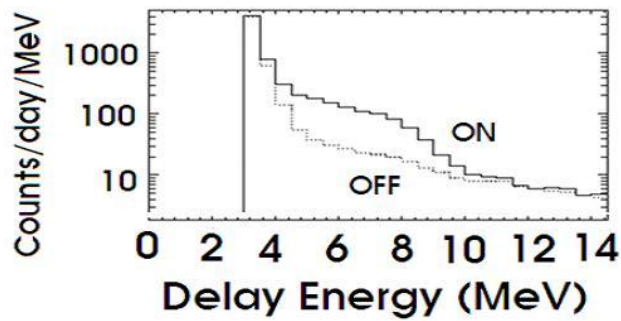

**Figure 7.** Spectrum recorded with a $m^3$ scintillation detector showing the antineutrinos detected and the signal to background ratio achieved in the vicinity of a power reactor (Ref. Bowden et al).

As the detector is moved farther from an anthropogenic source of antineutrinos, the signal to background ratio becomes unfavorable. The actual signal needs to be extracted from the observed spectrum from a detailed knowledge of all the sources contributing to the background signals, which mimic the signal of interest. Because of the unfavorable signal to background ratio at large distances from the source of fission antineutrinos, the actual signal need to be extracted from the observed spectrum from a detailed knowledge of all the sources contributing to the background signals which mimic the signal of interest in these experiments.

## BACKGROUND INTERFERENCES:

There are primarily four component sources in the background spectrum as shown in figure 8:
1. The intrinsic HPE contamination in the component materials used in the construction of the detector.
2. The cosmic-ray interactions, which produce signals similar to the signal of interest.
3. Other sources of interfering signals in the detector component materials.
4. Interference from fluctuating airborne radioactivity.

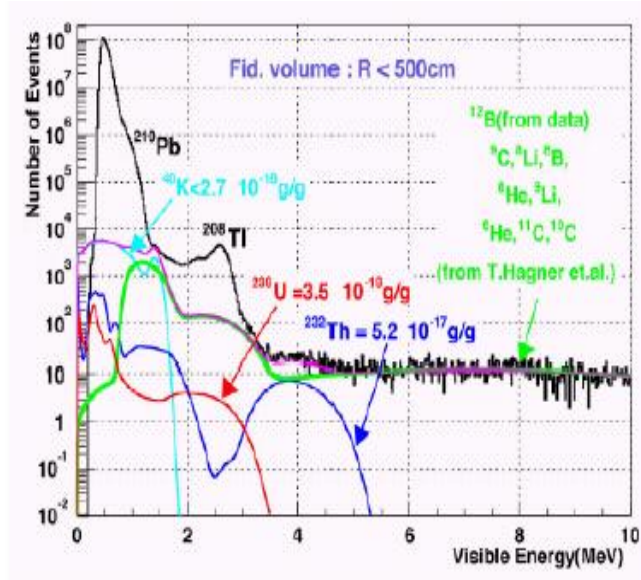

**Figure 8.** Interfering backgrounds as identified in the KamLAND detector. The calculated contribution from cosmogenic activity is shown in green. U and Th contributions are shown at their achieved levels of radiopurity, (Ref. KamLAND Collaboration: A. Kozlov).

# CONCLUSIONS

It can be seen from the above considerations that the research priorities must first be decided while considering the selection of a dedicated antineutrino detector. This will identify the design specifications for the antineutrino detector of choice. In this context, a policy decision is needed from the geoscience community. In addition, the cost and time considerations are of paramount importance in the research area of interest. It is clear that while a 1 meter cubed detector may be deployed at low cost in a short time to determine the total radioactivity of the Earth, a detector with tomographic and spectroscopic capabilities requires careful consideration regarding detection sensitivity and energy resolution. These additional considerations require the geosciences to invest considerable time, effort, manpower and capital. Experience shows that such large sophisticated detectors may take up to ten years to secure the needed funding and manpower, and another five years to construct and commission them. Only then can any data be expected from the dedicated detectors for geosciences.

For in depth understanding of the geophysics and geochemistry considerations, please refer to literature by authors such as Anderson, Fiorentini et al, Herndon, Learned and McDonough.

# ACKNOWLEDGEMENTS

This work is supported in part by funding from The NORM Group Organization, and Jagat Veda Peetha Foundation.